\documentclass{iau}
\usepackage{graphicx}

\usepackage{url}

\usepackage{xspace}

\def\apj{ApJ}

\def\aap{A\&A}

\def\mnras{MNRAS}

\newcommand{\isis}{\textsc{isis}\xspace}

\usepackage{natbib}
\usepackage{color,hyperref}
\hypersetup{breaklinks=true,colorlinks=true,
            linkcolor=blue,urlcolor=blue,anchorcolor=blue,citecolor=blue,filecolor=blue,
            bookmarks=true, bookmarksopen=true, bookmarksopenlevel=1}  

\title[Silicon K-edge Properties of Low-mass X-ray Binaries] 
{Silicon K-edge Dust Properties of \\ Neutron Star Low-mass X-ray Binaries}

\author[A. Danehkar]   
{A. Danehkar}

\affiliation{Department of Astronomy, University of Michigan, Ann Arbor, MI 48109, USA}

\pubyear{2023}
\volume{363}  
\pagerange{342-344}
\doi{\href{https://doi.org/10.1017/S174392132200045X}{10.1017/S174392132200045X}}
\jname{Neutron Star Astrophysics at the Crossroads: \\Magnetars and the Multimessenger Revolution}
\editors{E. Troja \& M. G. Baring, eds.}
\begin{document}

\maketitle

\begin{abstract}
The dust properties of the line-of-sight materials in neutron star low-mass X-ray binaries (LMXBs) can be probed by X-ray observations and laboratory experiments. We use a Markov chain Monte Carlo (MCMC) method to conduct a spectral analysis of \textit{Chandra} ACIS-S/HETG archival data of a sample of LMXBs, including GX\,5-1 and GX\,13+1. Our MCMC-based analysis puts constraints on the Si K-edge dust properties of the outflowing disk winds in this sample. Further X-ray observations of other LMXBs will help us better understand the grain features of dense outflows and accretion flows in neutron star binary systems.
\end{abstract}

The properties of silicate-based dusty materials in dense accretion disk and outflowing winds of low-mass X-ray binaries (LMXBs), such as grain and metallic content, can be determined from the K-edge absorption profile around 1.84 keV in the rest frame \citep{Schulz2016,Rogantini2020}.
It has been found that photoelectric edges in the X-ray soft band can be utilized to explore the common features of dust grains of the interstellar medium \citep[see e.g.,][]{Lee2005,Lee2009}. In particular, silicon K$\alpha$ edges might disclose the accreting properties of LMXBs, which allow us to assess better binary neutron star candidates for mergers. 

\begin{figure}[hbt!]
\centering
\includegraphics[width=0.4\textwidth, trim = 20 20 20 40, clip, angle=270]{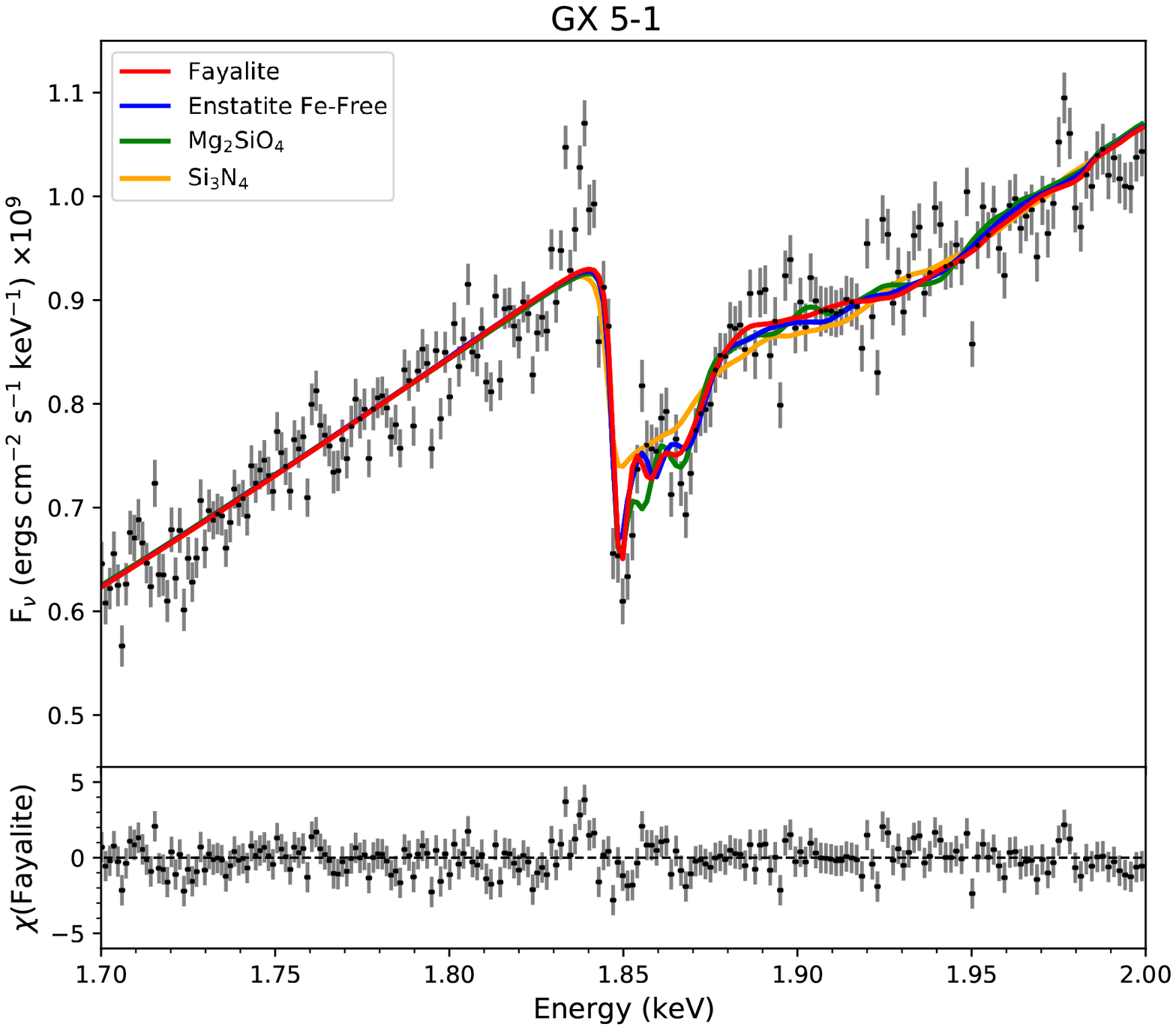}%
\includegraphics[width=0.4\textwidth, trim = 20 20 20 40, clip, angle=270]{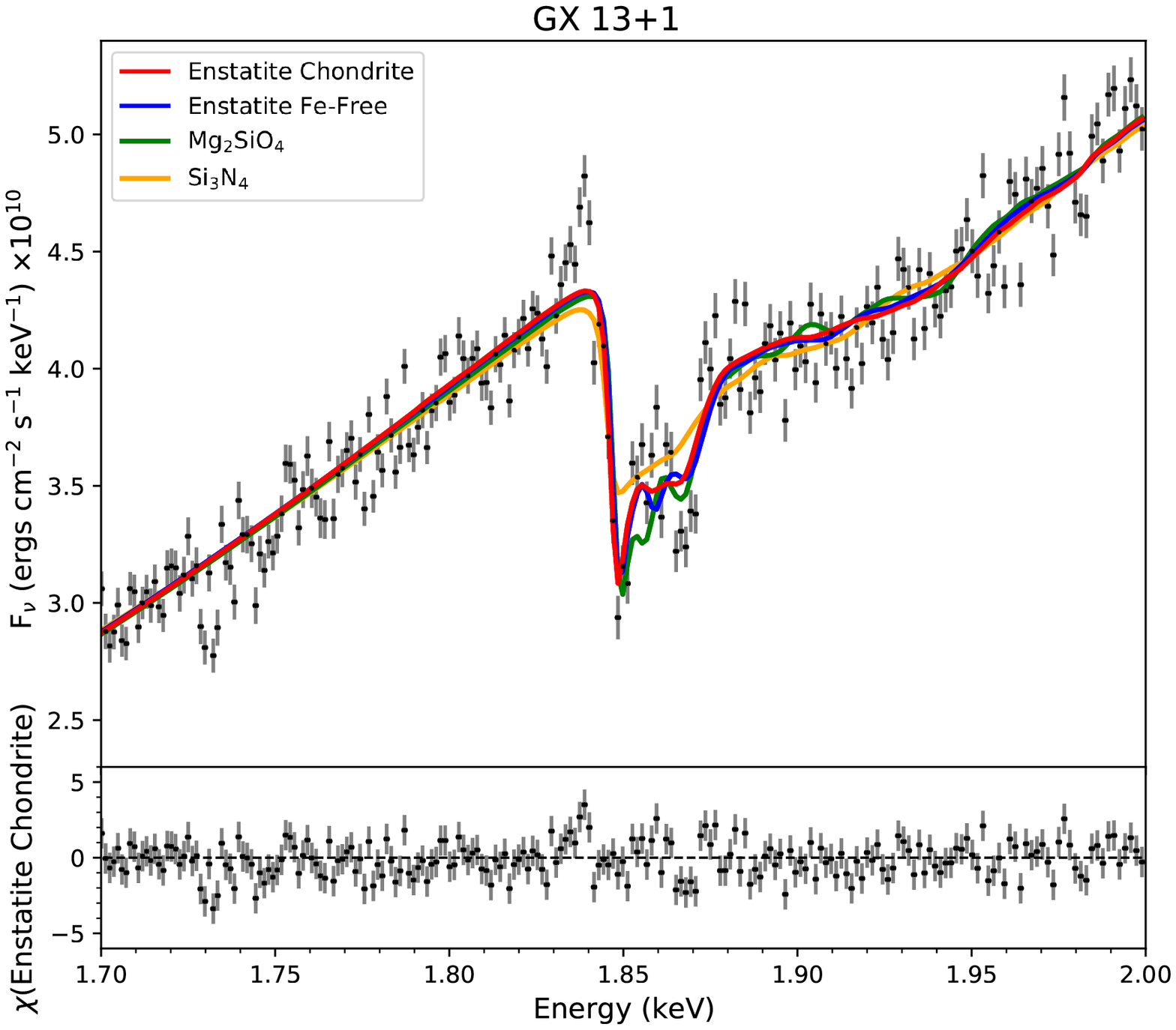}
\caption{Si K-edge absorption profiles of GX\,5-1 (left panel) and GX\,13+1 (right) in the combined MEG and HEG, ACIS-S/HETG \textit{Chandra} observations, fitted by various dust models, including Fayalite, Enstatite Chondrite, Enstatite Fe-Free, Mg$_2$SiO$_4$, and Si$_3$N$_4$.
}
\label{fig1}
\end{figure}

We considered \textit{Chandra} ACIS-S/HETG observations of a sample of LMXBs (GX\,5-1 and GX\,13+1 here). 
Cross-section data from various dust grains measured in laboratory experiments (shared by J.\,C. Lee) were fitted into the Si K-edge features in the \textit{Chandra} ACIS-S/HETG observations to constrain the dust composition of LMXBs (see Fig.\,\ref{fig1}).
The cross-sections include Fayalite, Enstatite Chondrite,
Enstatite Fe-Free, Mg$_2$SiO$_4$, and Si$_3$N$_4$. 
Spectral MCMC modeling was performed with the S-Lang \textsf{emcee} module (developed by M.\,A. Nowak, 2017) in the Interactive Spectral Interpretation System \citep[\isis;][]{Houck2000}.
In particular, MCMC-based fitting methods have been found to be promising in determining the best-fitting model parameters of dense media around compact objects in X-ray astrophysics \citep[e.g.,][]{Danehkar2018,Danehkar2021}.

\begin{figure}[hbt!]
\centering
\includegraphics[width=0.42\textwidth, trim = 20 15 15 0, clip, angle=0]{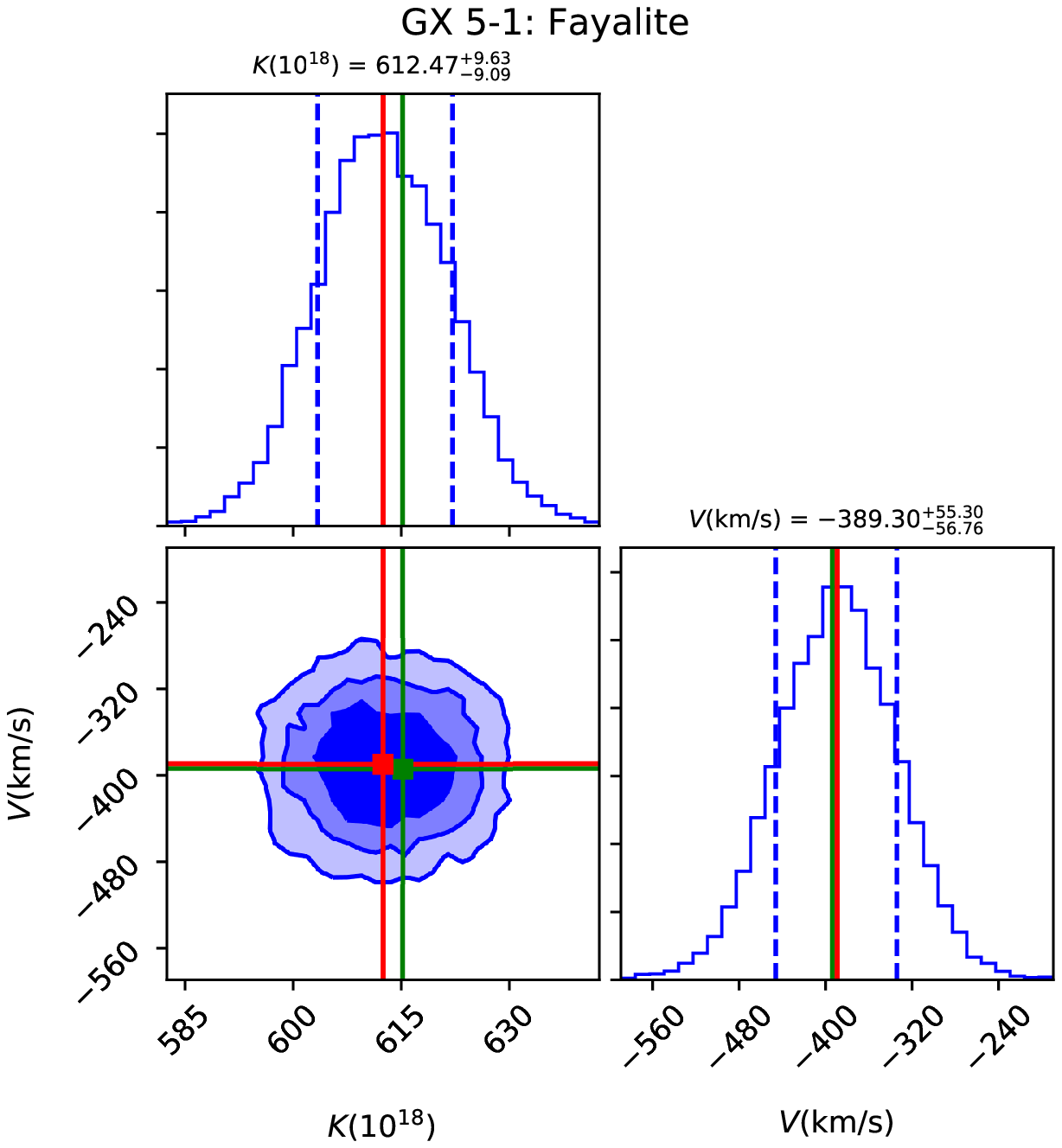}%
\includegraphics[width=0.42\textwidth, trim = 20 15 15 0, clip, angle=0]{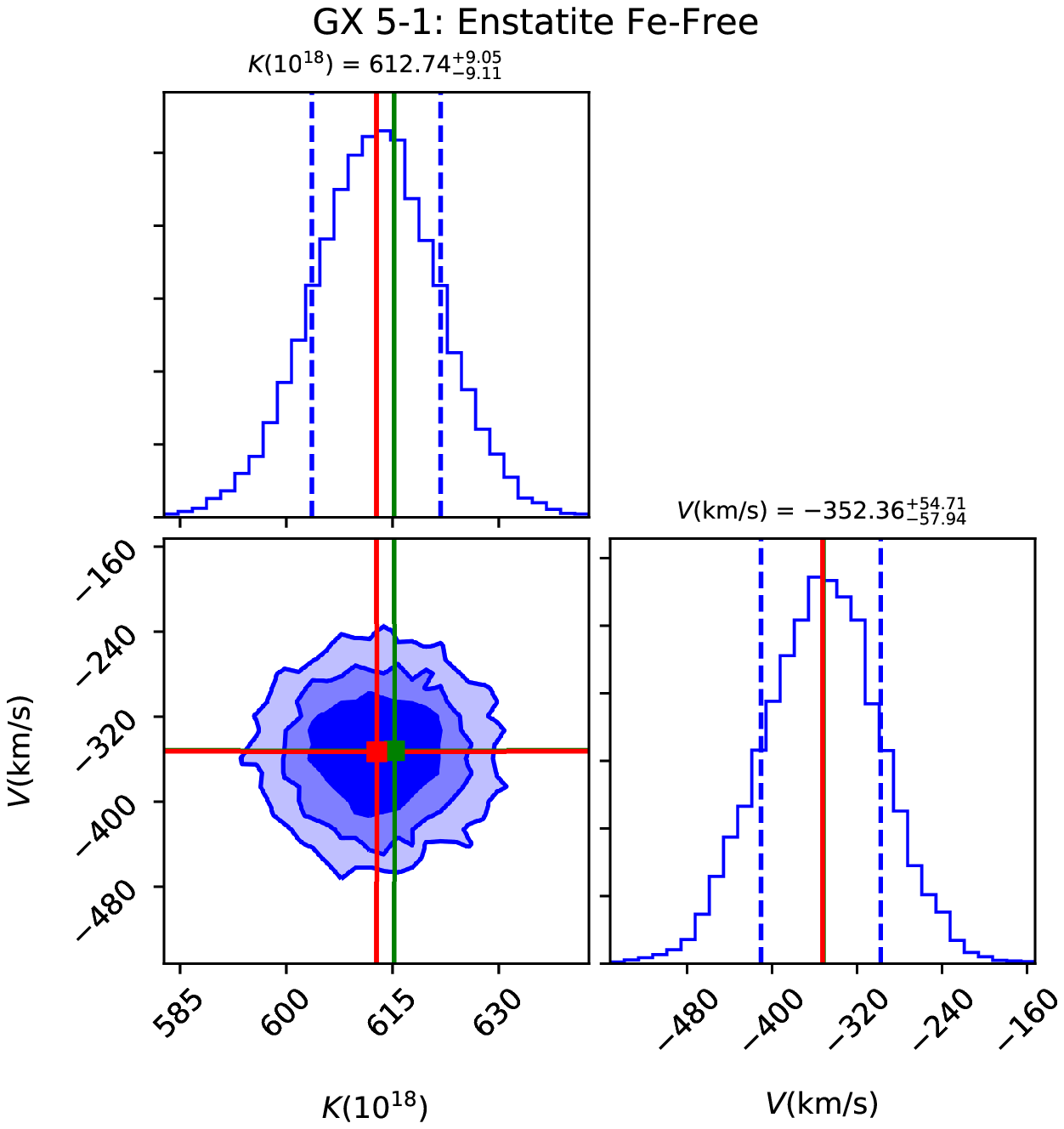}\\
\includegraphics[width=0.42\textwidth, trim = 20 15 15 0, clip, angle=0]{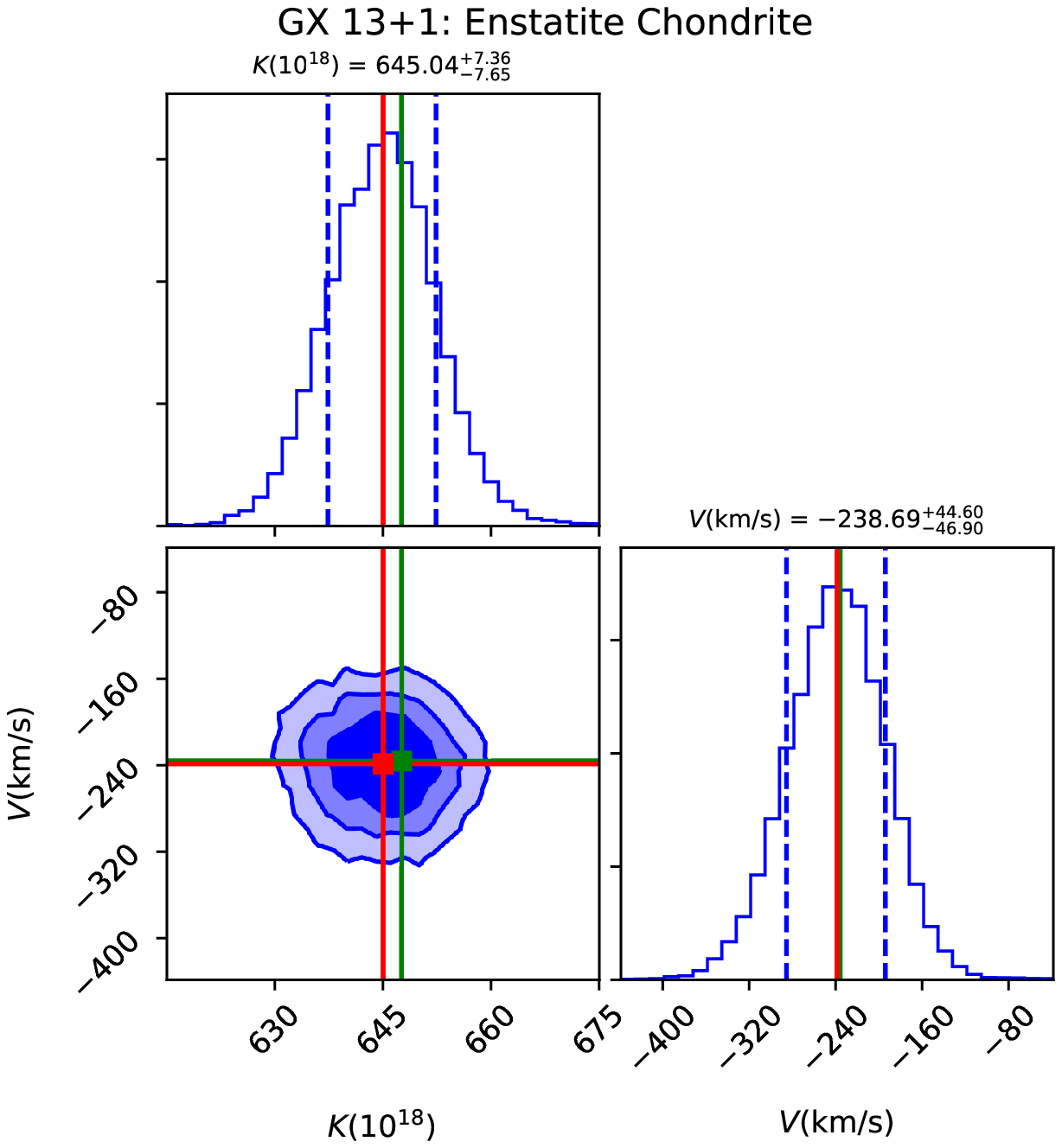}%
\includegraphics[width=0.42\textwidth, trim = 20 15 15 0, clip, angle=0]{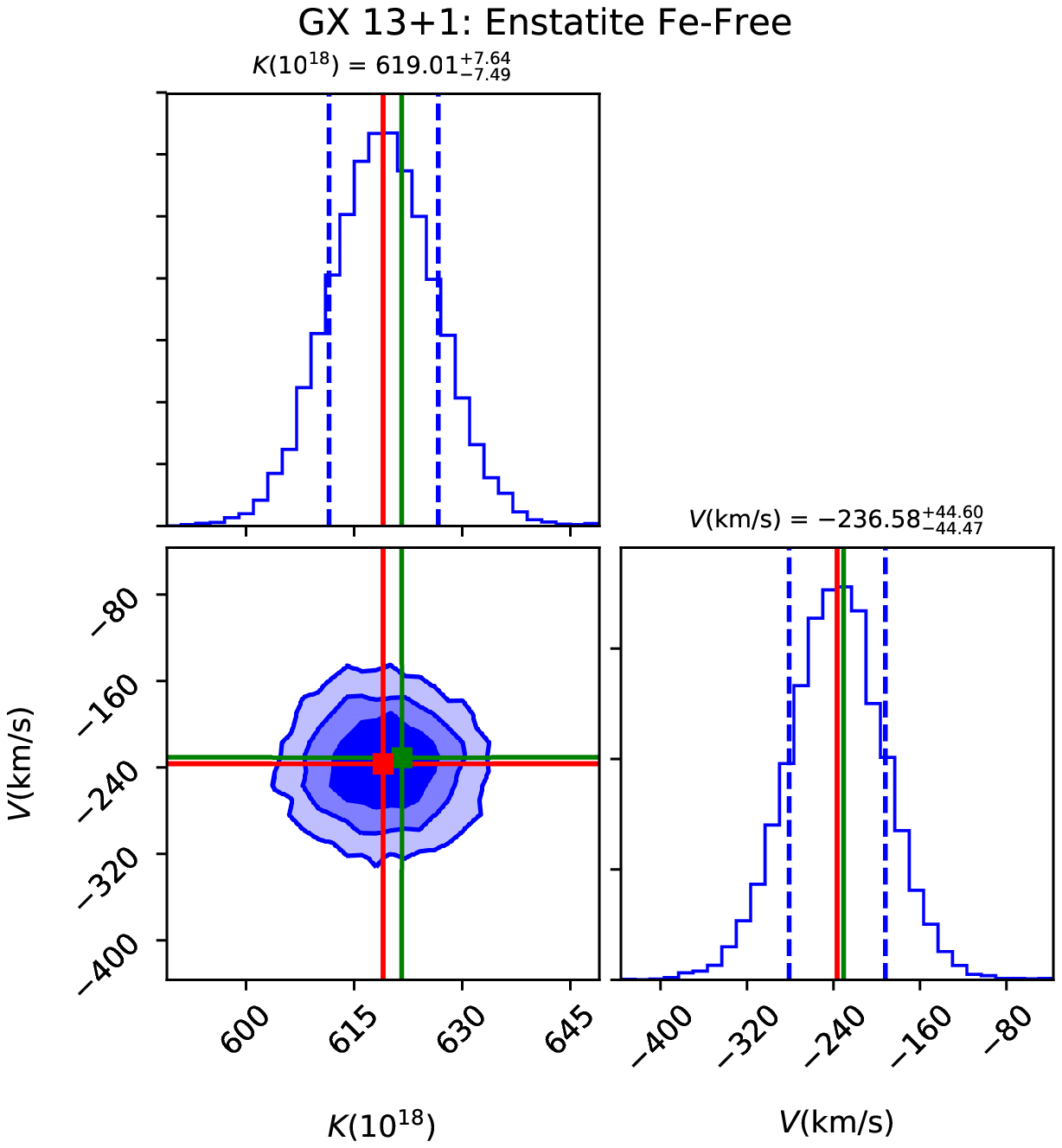}
\caption{Posterior probability distributions of the normalization factor $K$ and outflow velocity $V$
for the dust models fitted to Si K-edge absorption features of GX\,5-1 (top panels; Fayalite and Enstatite Fe-Free) and GX\,13+1 (bottom; Enstatite Chondrite and Enstatite Fe-Free).
}
\label{fig2}
\end{figure}

Our MCMC analysis suggests that the lines-of-sight dusty outflowing materials in GX\,5-1 and GX\,13+1 (see Fig.\,\ref{fig2}) are well fitted with Fayalite, Enstatite Chondrite, and Enstatite Fe-Free. We also get acceptable fits with
Mg$_2$SiO$_4$. However, the Si$_3$N$_4$ model is unable to well reproduce the Si K-edge absorption characteristics in these LMXBs.
As seen in Fig.\,\ref{fig2}, the cross-section models of Fayalite and Enstatite Fe-Free fitted to 
the Si K-edge absorption profile of GX\,5-1 correspond to blueshifted outflow velocities of $\approx -390$ and $-350$ km\,s$^{-1}$, respectively. 
Similarly, an MCMC-based spectral analysis of the silicon K$\alpha$ edge in GX\,13+1 yields 
$\approx -240$ km\,s$^{-1}$ with the Enstatite Chondrite and Enstatite Fe-Free cross-sections, while a mean K$\alpha$ outflow velocity of $\approx -400$ km\,s$^{-1}$ was found by \citet{Allen2018}. 
The blueshifted Si K-edge absorption features could be associated with outflowing disk winds along the line of sight in these  LMXBs.


\end{document}